Digital Normativity: A challenge for human subjectivization and free will

Eric Fourneret[1,2] and Blaise Yvert[1,2]

1. Inserm, Braintech Lab, Grenoble, France
2. Univ Grenoble Alpes, BrainTech Lab, Grenoble France

## Abstract

Over the past decade, artificial intelligence has demonstrated its efficiency in many different applications and a huge number of algorithms have become central and ubiquitous in our life. Their growing interest is essentially based on their capability to synthesize and process large amounts of data, and to help humans making decisions in a world of increasing complexity. Yet, the effectiveness of algorithms in bringing more and more relevant recommendations to humans may start to compete with human-alone decisions based on values other than pure efficacy. Here, we examine this tension in light of the emergence of several forms of digital normativity, and analyze how this normative role of AI may influence the ability of humans to remain subject of their life. The advent of AI technology imposes a need to achieve a balance between concrete material progress and progress of the mind to avoid any form of servitude. It has become essential that an ethical reflection accompany the current developments of intelligent algorithms beyond the sole question of their social acceptability. Such reflection should be anchored where AI technologies are being developed as well as in educational programs where their implications can be explained.

## Introduction

Over the past two decades, digital technology has deeply changed our way of life and our relation to the world and to others, especially given today's unprecedented computing and communication technologies and possibilities for easy access to large amounts of knowledge and information. Recent advances in artificial intelligence (AI) are now opening up ever-newer perspectives, enabling possible cooperation between humankind and digital technologies to think the world. This new step is possibly engendering a profound change in the course of human evolution, triggering both enthusiasm and concerns. Taking this turn, humans will need to learn how to live with the new possibilities that AI offers, so that new forms of "being-in-the-world" will develop for their benefit. In particular, they are invited to re-evaluate their role in societies merged with AI, where they will need to remain subjects of their collective and personal lives.

AI algorithms based on supervised learning are built by tuning the parameters of mathematical models to obtain the best possible correspondence between a set of input data and a set of output data. Once trained, they can then be used to make new predictions from new observations. Striking examples are artificial neural networks (ANNs) such as deep neural networks (DNNs)[1], which are inspired from biological neural networks. Thanks to the current availability of large computational power (e.g., GPUs or TPUs), AI algorithms can be trained on huge amounts of data over limited time spans, thereby definitely out-performing human learning times by several orders of magnitude. These algorithms can thus benefit in a blitz from the massive amount of available knowledge that mankind has accumulated over its history and/or over the world. Moreover, reinforcement learning[2] can also be a very efficient way to train ANNs on extremely large datasets self-generated in very limited time compared to the time humans would need to create them. As a result, algorithms can easily become more efficient than humans on specific tasks after relatively short training periods (typically a few hours or days, as compared to several years for humans). The past decade has demonstrated their technical efficiency in multiple specific domains, as for instance, optimization of financial transactions, speech or text recognition[3], language translation[4], real-time image content analysis, autonomous driving[5], or playing chess or go[6]. Beyond engineering applications, they are also starting to see use in medicine by assisting physicians in reaching diagnoses[7][8][9], or in the field of neuroprosthetics to design rehabilitation solutions that compensate functional deficits following a disease or an injury to the central nervous system[10][11][12][13]. AI algorithms also find applications in social contexts, such as providing assistance to judges with decisions rendered in courts of law[1][14]. They have also shown greater efficacy than humans in inferring the sexual orientation of an individual based on only a few images [15]. This multiplicity of technical demonstrations is thus progressively bringing AI central and ubiquitous in human life.

The growing role of algorithms is essentially based on their increasing efficiency to synthesize and process large amounts of data, helping humans to make decisions in a world of increasing complexity. Yet, the effectiveness of algorithms in bringing more and more relevant recommendations to humans may start to compete with human-alone decisions based on values other than pure efficacy. Here, we examine this tension in light of the emergence of several forms of digital normativity, and analyze how this normative role of AI may influence the ability of humans to remain subject of their life.

## 1. The advent of digital normativity in multiple forms

---

[1] Public Safety Assessment by Laura and John Arnold Foundation, arnoldfoundation.org ; Kennedy, House, Williams, 2013

The increasing role of AI algorithms has been accompanied by the advent of a digital normativity that can take several forms. The first one comes from the fact that algorithms tend to reproduce the trends that are most present in the data on which they have been trained. This creates a normalized view of the problem they are intended to solve. Interestingly enough for humans, the level of details that algorithms might be able to discriminate can be high, as for instance, in automated image pattern recognition[16] or autonomous driving[17]. This first form of digital normativity may thus often be satisfying enough for humans for them to rely on algorithmic recommendations. However, the automatic and thus objective processing of large datasets restitute general trends present in these datasets, whether ethically good or bad[18].

Another form of digital normativity arises from the use of predictive algorithms trained on objective observational data without accounting for the course through which this data has been generated. For instance, predictive algorithms used to provide a customer with purchasing suggestions only rely on previous purchases made by the same and other customers, without access to the personal reasons that led them to these purchases. This form of automatic data processing thus eliminates the inherent subjectivity of the customer. The individual is objectivized (normalized) by the algorithm[19]. This second form of digital normativity is actually a recursive and dynamic process: algorithmic recommendations emanating from previous human actions in turn influence their next actions[20][21].

The normative role of algorithms takes a third form when their efficiency outperforms that of humans. If, for a given application, an algorithm has a higher predictive power than by any human expert, it may become reasonable to rely solely on this algorithm for making decisions. In this manner, the algorithm creates the norm by imposing its efficacy. The efficacy becoming the norm, the question becomes whether the role of humans in determining for themselves the finality of this efficiency could be challenged.

2. **The human subject in the context of digital normativity**

The notion of "subject" is central in philosophy[22] (Descartes takes the "*cogito ergo sum*", "I think", as a starting point of all his philosophical researches). Being subject is the result of a construction (process of subjectivization), leading one to be aware of oneself as a subject, i.e., to be free and responsible for one's actions, and to be at the foundation of one's representations and judgments: "The *I think* must be able to accompany all my representations"[23]. This capacity to be an actor and be responsible for one's life is progressively acquired as the result of situations encountered during a person's life experience, including education, professional life, and more. A subject thus gets built according to his or her environment, which encompass the technologies that are available and their evolution[24]. Subjectivization is thus a fundamental movement that brings an individual to a state of subject[25], and the question becomes whether this movement can be affected in the context of digital normativity. Can AI technology weaken, or on the contrary, be a new way to boost the capacity of human individuals to build themselves, to become subjects of their lives, to become authors of the norms that organize their individual and collective lives?

Until recently, there would have been no valid reason for raising such a question. Indeed, it is humans who develop artificial intelligence algorithms and are thus the ones who impose the rules by which these algorithms learn to make predictions. This role places humans as the instructors of machines. In other words, human action remains required, and if algorithms make autonomous decisions without human intervention, these decisions still rely on a set of rules established by humans. In this sense, the autonomy of algorithms can be seen only in a technical sense, namely as an autonomy of execution, which preserves human subjectivization. However, algorithms are becoming ubiquitous in many aspects of the lives of people, and may thus also play a role in their construction as subjects. Indeed, subjectivization is progressively acquired by learning through the experiences acquired during one's life (e.g., education, social relationships, professional experiences, sports, entertainment, etc.). Hence, because AI algorithms now constitute an important part of the human environment, they participate in the process of subjectivization. On the one hand, they offer unprecedented opportunities to strengthen this process. For instance, a search engine gives access to a huge amount of available knowledge in just a few clicks, thus offering new routes to any individual to build his or her critical judgment, a mandatory ability to become a subject. On the other hand, they may also bias the process of subjectivization. This could, for instance, be the case of a search engine that determines which results are put at the top of the list based on a statistical inference not accounting for the individual as a subject, but rather on raw desubjectivized data.

Once subjectivization has been acquired, AI algorithms may also further influence how a person exerts him or herself as subject. Expressing oneself as the subject of actions may indeed no longer be desired by humans whenever algorithms may efficiently handle for them the tasks to be accomplished. This could be for sake of physical comfort when a task is physically demanding (e.g., requiring focused attention over a long period of time, such as driving long distances), or for the sake of psychological comfort when a task or a decision engages a moral responsibility that is difficult to endorse. For instance, algorithms are used in the justice system in Belgium to evaluate the risk of recidivism and help determine whether an imprisoned individual should benefit from anticipatory freedom. The responsibility of the judges may be increased and more difficult to stand if they decide against the recommendation of an algorithm. Indeed, in such case, if their decision is found to be inappropriate in the future, they could be opposed to have acted against an algorithmic decision considered more objective than any human decision[26]. Although it remains theoretically possible to resist this normative role of AI algorithms, the associated amplification of human responsibility could become so much of a deterrent that disobedience would no longer be possible in practice. The ubiquity of algorithms in our daily lives may thus offer multiple opportunities for humans to progressively disengage from their role of subjects of their lives[27]. Therefore, attaining a stage where humans would systematically have to comply with the predictions of algorithms is no longer impossible, at least in specific domains. Algorithms could then become prescriptive by imposing their digital normativity, and thus no longer restricted with an autonomy of execution but acquiring a form of moral autonomy at the expense of human moral autonomy, and hence subjectivization. This could result in the emergence of a form of governance without subject.

### 3. The risk of a silent human desubjectivization

Despite the importance taken by algorithms in the organization of human societies, algorithms do not decide alone and, even if their complexity is highly evolved, humans still keep control over their own decisions as they decide how algorithms are built and how and for which purposes they are used. In other words, a cooperative relationship between humans and AI exists: on the one hand a form of expertise (the algorithm) and on the other hand the power to decide (humans). Each needs the other but both do not merge as one. Indeed, a qualification of decision differs from a qualification of expertise: a power to decide can be exerted in absence of expertise or against the recommendation of an expert, and conversely, an expert is not necessarily legitimate to decide. A decision requires intuition and know-how more than only argued and demonstrated knowledge, even though knowledge is a necessary, but not sufficient, step in the process of decision making[28][29][30]. Deciding is acting with doubts, thus accepting the risk of making errors, which is why any decision differs from simply knowledge. However, if humans were to decide to dispossess themselves from their power of decision, they would abandon an essential feature of their humanity, namely their capability to learn from errors, and would consequently put into question their power of perfectibility[31].

For current generations, humans remain vigilant regarding the power conferred to algorithms. But what about future generations born after the emergence of digital normativity, and thus well habituated with its ubiquity? When introducing the notion of voluntary servitude, La Boétie already seized this question to understand the foundations of despotic political powers. He pointed out that in a process of oppression, people are at first aware of losing their freedom but that the next generations make this world of oppression the rule and become either unaware of their servitude or accustomed to it: "(...) Those who come after serve without regret, and willingly do what their predecessors had done by constraint"[32]. The advent of algorithmic governmentality is however unique in the sense that it does not impose itself by any brutal or violent physical or moral means. By contrast, it progressively meddles into our environment through different changes of practice and stabilizes always by habitude. This is where a risk of silent human desubjectivization could take root.

This risk is then further reinforced by the challenge of explicability of algorithms. A predictive algorithm based on artificial neural networks is composed of mathematical functions governed by a set of parameters. These functions are chained together according to a specified architecture (defined by the developer of the algorithm) so that the outputs of some of them become the inputs of others. The parameters of the different functions are then optimized to make the algorithm most efficient on the task it is intended to solve. Although the methods used to optimize the parameters of AI algorithms are well understood, the resulting set of optimal parameters does not generally represent any intuitive or ecological meaning for a human being. In other words, these parameters may be optimal to best map observations onto predictions, but cannot be explained in any form of intelligibility comprehensible by humans. Specific studies aim at exploring and understanding the structure of trained artificial neural networks[33] but, most of the time, these algorithms remain opaque, with the impossibility for humans to fully understand how input observations are processed to infer predictions. Relying on algorithmic decisions may thus become critical when humans can no longer access the reasoning allowing the expertise of algorithms to surpass their own expertise. Indeed, understanding reasoning hinders applying a decision blindly without critical evaluation, and thus silencing the capability of the human subject to distinguish between the fair and the unfair.

## Conclusions

AI has clearly become a unique opportunity for accompanying the evolution of human well-being, especially by freeing people from multiple constraints imposed by nature and helping in their quest for a better life. The ubiquity of AI algorithms in our everyday lives however brings a new major ethical challenge for humans: to preserve our capability to remain subjects and not only agents. Indeed, the advent of AI technology imposes a need to achieve a balance between concrete material progress and progress of the mind. AI should remain a tool, and the finality of this tool should remain in the hands of humans for the sake of freedom. Ethics should help in this respect by avoiding both "technophoby" and "technophily", with the only constant concern being that what humans gain from technology is not at the expense of freedom and free will. The solution is not to hamper the development of algorithms, but to ensure that it does not lead humans to lose their capability to make choices with respect to human dignity and the capability to undo what has been made. Indeed, humans need to define and change themselves the rules by which they organize their collective life, including their life with AI technology. Far from either completely embracing or completely rejecting AI technologies, it has become essential that an ethical reflection accompany the current developments of intelligent algorithms beyond the sole question of their social acceptability. A thoughtful ethical reflection cannot be conducted independently from the scientific actors of AI technology, and needs to be anchored with respect to the concrete applications that AI addresses, apprehending their values and their aims. Moreover, it should also aim at educating the next generations about the ethical implications of AI. This double scientific and societal anchoring of a pragmatic ethic is mandatory to preserve human subjectivization, free will, and freedom in the long term: "Techniques always bring with them the world in which they will make sense"[34]. AI should not be developed to invent the future for us, but rather to help us invent our future.

## Acknowledgments

This work was supported by the European Union's Horizon 2020 research and innovation program under Grant Agreement No. 732032 (BrainCom), and by the French National Research Agency under Grant Agreement No. ANR-16-CE19-0005-01 (Brainspeak).

Clinical Implementation," *Radiology*, p. 180694, 2018.